\documentclass[a4paper,11pt]{article}
\pdfoutput=1  
\usepackage[utf8]{inputenc}
\usepackage{jheppub}

\usepackage{amssymb}
\usepackage{physics}
\usepackage{graphicx}
\usepackage{hyperref}
\usepackage{slashed}
\usepackage{bbm}

\newcommand{\be}{\begin{equation}}
\newcommand{\ee}{\end{equation}}
\newcommand{\bea}{\begin{eqnarray}}
\newcommand{\eea}{\end{eqnarray}}

\usepackage{color}
\usepackage[normalem]{ulem}

\def \del{\partial}
\def\bea{\begin{eqnarray}}
\def\eea{\end{eqnarray}}
\newcommand{\nn}{\nonumber}

\begin{document}
 \title{
Are there ALPs in the 
 asymptotically safe
landscape?
 }
 \author[a]{Gustavo P. de Brito,}
 \author[a]{Astrid Eichhorn,}
 \author[a]{Rafael R. Lino dos Santos}

\affiliation[a]{CP3-Origins,  University  of  Southern  Denmark,  Campusvej  55,  DK-5230  Odense  M,  Denmark}

\emailAdd{gustavo@cp3.sdu.dk}
\emailAdd{eichhorn@cp3.sdu.dk}
\emailAdd{rado@cp3.sdu.dk}

\abstract{We investigate axion-like particles (ALPs) in the context of asymptotically safe gravity-matter systems. The ALP-photon interaction, which facilitates experimental searches for ALPs, is a dimension-5-operator. Quantum fluctuations of gravity lower its scaling dimension, and the ALP-photon coupling can become asymptotically free or even asymptotically safe. However, quantum fluctuations of gravity need to be strong to overcome the canonical scaling and this strong-gravity regime is in tension with the weak-gravity bound in asymptotic safety. Thus, we tentatively conclude that fundamental ALPs can likely not be accommodated in asymptotically safe gravity-matter systems. In turn, an experimental discovery of an ALP would thus shed valuable light on the quantum nature of gravity.} 

\maketitle

\section{Introduction}\label{sec:Intro}

We explore whether axion-like particles (ALPs) are compatible with asymptotically safe quantum gravity. On the one hand, this is of interest for ALP phenomenology, as it could provide theoretical constraints on the ALP parameter space. On the other hand, this is of interest for quantum gravity, because it could give rise to observational tests of asymptotically safe quantum gravity.\\

ALPs are dark-matter candidates, see \cite{Arias:2012az,Ringwald:2012hr,Essig:2013lka,Ringwald:2014vqa,Irastorza:2018dyq, Choi:2020rgn,Irastorza:2021tdu} for reviews. They are typically explored within an effective-field theory setting, where a huge parameter space, spanned by the ALP-photon coupling and the ALP-mass, is subject to experimental scrutiny. We instead explore ALPs within the context of an ultraviolet complete framework of all fundamental interactions, including gravity, the asymptotic safety framework. In short, asymptotic safety relies on the dynamical emergence of a novel symmetry, quantum scale symmetry, at high energies. In \cite{Shaposhnikov:2009pv,Harst:2011zx,Eichhorn:2017ylw,Eichhorn:2017lry,Eichhorn:2018whv}, it has been shown that the extra symmetry -- much like any other symmetry that is imposed on the interactions of a given set of fields -- constrains the allowed values of couplings. 
Thus we expect that the embedding of ALPs into such an ultraviolet complete setting provides novel theoretical constraints on the ALP parameter space -- which might even be strong enough to rule out the entire ALP parameter space. In this case, the landscape of beyond-Standard-Model settings that are compatible with asymptotic safety would not contain any models that feature ALPs. Thus, we ask \textit{whether or not there are ALPs in the asymptotically safe landscape}.

More specifically, ALPs are ultra-light candidates for dark matter. Ultra-light dark matter (ULDM) candidates, see \cite{Bernal:2017kxu,Ferreira:2020fam} for reviews,  have masses in the range $10^{-24} \,\textmd{eV} < m < 1 \,\textmd{eV}$. 
Because they are lighter than $1 \,\rm{keV}$, ULDM candidates can only be bosons and cannot be thermal relics. Instead, their production mechanism is non-thermal.  For axions, the misalignment mechanism is an example for a non-thermal production process \cite{Preskill:1982cy,Abbott:1982af,Dine:1982ah}, see  \cite{Sikivie:2006ni, Marsh:2015xka}, for reviews on axion cosmology.  

In our paper, we explicitly focus on ALPs, not on the QCD axion. The reason is that we consider a scenario where a pseudoscalar ALP corresponds to a fundamental field, whereas the QCD axion is a pseudo-Nambu-Goldstone boson, with a more extensive set of fundamental fields, see \cite{Peccei:1977hh,Peccei:1977ur, Weinberg:1977ma,Wilczek:1977pj}. The QCD axion can solve the strong CP problem, which is a fine-tuning problem.  If one takes the view that fine-tuning might well be much more natural than it has commonly been viewed, this motivation to introduce an axion in QCD is absent. This view of fine-tuning is well-justified by the observation of the Higgs and so far no other new physics at the LHC, as well as the value of the cosmological constant. Then, ALPs remain attractive dark-matter candidates, offering an alternative to the WIMP-scenario, for which experiments have already excluded substantial parts of parameter space \cite{Schumann:2019eaa}.

Experimental searches for ALPs, see \cite{Graham:2015ouw,Irastorza:2018dyq,Carena:2788559} for reviews, rely on the Primakoff process, i.e., the ALP-photon conversion \cite{Raffelt:1987im,Sikivie:1983ip,PhysRevD.18.1829}. This enables both light-shining-through-wall experiments \cite{Ruoso:1992nx,Cameron:1993mr,Fouche:2008jk,OSQAR:2007oyv,GammeVT-969:2007pci,Redondo:2010dp,Ehret:2010mh,Bahre:2013ywa,Betz:2013dza,OSQAR:2015qdv} as well as helioscopes \cite{SOLAX:1997lpz,Moriyama:1998kd,COSME:2001jci,Inoue:2002qy,Inoue:2008zp, Armengaud:2014gea, CAST:2017uph, IAXO:2019mpb}, which search for solar axions. Additionally, haloscopes \cite{PhysRevLett.104.041301, Caldwell:2016dcw, McAllister:2017lkb, Alesini:2017ifp, ADMX:2018gho, Ouellet:2018beu,HAYSTAC:2018rwy,Melcon:2018dba} detect axions from the galactic dark matter halo, typically using microwave cavities. Other dark-matter searches have also searched for ALPs \cite{CDMS:2009fba,Armengaud:2013rta,XENON100:2014csq,LUX:2017glr}. 
Furthermore, cosmological \cite{Marsh:2015xka} and astrophysical \cite{Raffelt:2006cw} constraints for ALPs exist, e.g., from stability of massive stars \cite{Friedland:2012hj,Ayala:2014pea} and supernova observations \cite{Payez:2014xsa}.\\

From an experimental point of view, probing the entire parameter space spanned by the ALP-photon coupling $g$ and the ALP mass $m$ is a challenge.
From a theoretical point of view, the perturbative non-renormalizability of the dimension-5  ALP-photon  operator $\phi\, F_{\mu\nu}\widetilde{F}^{\mu\nu}$  (with the ALP $\phi$, the Abelian field strength tensor $F_{\mu\nu}$ and its dual $\widetilde{F}^{\mu\nu}$) is a challenge that might be solvable in a non-perturbative way \cite{Eichhorn:2012uv}, but only at asymptotically high scales, where quantum gravity cannot be neglected. Thus, we explore whether both challenges can be solved in a theory which contains quantum fluctuations of gravity at very high energies.  Quantum gravity fluctuations are dynamically negligible at the energy scales at which experiments operate. 

However, first these fluctuations could solve the problem of perturbative renormalizability, if the ALP-photon coupling becomes asymptotically safe or asymptotically free under the impact of gravity. 
This may be possible because quantum gravitational fluctuations effectively act akin to a change in dimensionality and thus change the scaling dimensions of couplings. We explore this question in a toy model, where we neglect all fields of the Standard-Model except the Abelian gauge field. Because all other fields are absent, there is no spontaneous symmetry breaking and the $U(1)$ symmetry at high and low energies is the same, unlike in the Standard-Model, where the $SU(2)\times U(1)_Y$ is broken to $U(1)_{\rm em}$.

Second, requiring quantum scale symmetry, i.e., asymptotic safety, at high energies, often determines the values of couplings at all scales. If this mechanism is realized for the ALP-photon coupling, then the asymptotic-safety requirement would reduce the theoretically viable parameter space for these models. 
Thus, experimental results, including, e.g., an update by the XENON-collaboration \cite{XENON:2020rca}, could shed valuable light on the quantum nature of gravity.\\

This paper is structured as follows: In Sec.~\ref{sec:ASQGmatt} we introduce asymptotically safe gravity-matter systems, their predictive power and the techniques that enable us to search for asymptotic safety in gravity-matter systems. In particular, in Sec.~\ref{sec:WGB}, we introduce the weak-gravity bound, which is an important bound on gravity-matter systems in asymptotic safety and will play a crucial role in understanding ALPs in asymptotic safety. In Sec.~\ref{sec:Results} we introduce the setup of our calculation and the underlying assumptions, before then reporting on the results. We conclude in Sec.~\ref{sec:conclusions}, where we also offer an outlook.  In the appendices, we give additional details on our calculation.

\section{Asymptotically safe gravity matter systems}\label{sec:ASQGmatt}

In this work, we explore the interplay between asymptotically safe quantum gravity and matter in order to quantify the \textit{predictive power of asymptotic safety} for ALPs. In this section, we introduce the asymptotically safe ``landscape", i.e., the matter models compatible with an embedding into asymptotic safety, the functional Renormalization Group (FRG) machinery, and the interplay between matter and gravity.

\subsection{The asymptotic safety paradigm and its predictive power}
Asymptotic safety \cite{2010grae.book.....H} provides an ultraviolet (UV) completion for effective quantum field theories by realizing quantum scale symmetry \cite{Eichhorn:2018yfc,Wetterich:2019qzx}. Generically, quantum fluctuations break scale symmetry and lead to running, scale-dependent couplings. However, for special values of interaction strengths, it is possible to balance out quantum fluctuations and achieve scale symmetry at finite interactions. 
In a Renormalization Group (RG) language, asymptotic safety is equivalent to an interacting fixed point of the RG flow.
Therefore, asymptotic safety generalizes the concept of asymptotic freedom, associated with a free UV fixed point. Asymptotic safety is known to be realized in a number of different quantum field theories, even including Standard-Model-like settings in four dimensions, see the review \cite{Eichhorn:2018yfc}  for an overview. 

Asymptotic safety is a particularly attractive paradigm for quantum theories including gravity, because it could provide a quantum field theoretic understanding of gravity.
General Relativity is only compatible with standard perturbative quantization up to the Planck scale \cite{Donoghue:2012zc}, implying a loss of predictivity. An asymptotically safe fixed point is expected to restore predictivity. 
Following the seminal work by Reuter \cite{Reuter:1996cp}, compelling evidence for asymptotically safe quantum gravity is mounting \cite{Souma:1999at,Lauscher:2001ya,Reuter:2001ag,Lauscher:2002sq,Litim:2003vp,Codello:2006in,Machado:2007ea,Codello:2008vh,Benedetti:2009rx,Eichhorn:2009ah,Manrique:2010am,Eichhorn:2010tb,Groh:2010ta,Dietz:2012ic,Christiansen:2012rx,Rechenberger:2012pm,Falls:2013bv,Ohta:2013uca,Eichhorn:2013xr,Falls:2014tra,Codello:2013fpa,Christiansen:2014raa,Demmel:2015oqa,Gies:2015tca,Christiansen:2015rva,Ohta:2015fcu,Ohta:2015efa,Falls:2015qga,Eichhorn:2015bna,Gies:2016con,Denz:2016qks,Biemans:2016rvp,Falls:2016msz,Falls:2016wsa,deAlwis:2017ysy,Christiansen:2017bsy,Falls:2017lst,Houthoff:2017oam,Falls:2017cze,Becker:2017tcx,Knorr:2017fus,Knorr:2017mhu,DeBrito:2018hur,Eichhorn:2018ydy,Falls:2018ylp,Bosma:2019aiu,Knorr:2019atm,Falls:2020qhj,Kluth:2020bdv,Knorr:2021slg,Bonanno:2021squ,Baldazzi:2021orb,Sen:2021ffc,Mitchell:2021qjr,Knorr:2021iwv,Baldazzi:2021fye,Fehre:2021eob},
for reviews, see for instance \cite{Eichhorn:2017egq,Percacci:2017fkn, Reuter:2019byg,Pereira:2019dbn,Reichert:2020mja,Pawlowski:2020qer}. A critical discussion of the state-of-the-art in asymptotically safe quantum gravity can be found in \cite{Bonanno:2020bil}. Moreover, there are searches for asymptotic safety using other complementary approaches. For instance, in lattice techniques, asymptotic safety manifests as a second-order phase transition \cite{Loll:2019rdj}; there is also evidence for asymptotic safety coming from a background-independent setting defined by tensor-models \cite{Eichhorn:2018phj,Eichhorn:2019hsa}. \\

Asymptotic safety corresponds to an interacting fixed point, which is defined by the condition: 
\begin{eqnarray}
	\beta_{g_i}\mid_{g_i=g_{i\, \ast}}\equiv k \del_k\, g_i (k)\mid_{g_i=g_{i\, \ast}}=0, 
\end{eqnarray}
where $\beta_{g_i}$ is the beta function associated with the dimensionless coupling $g_i$ and $i$ labels all possible interaction structures in the theory. For each running coupling $ \bar{g}_i =  \bar{g}_i  (k)$ of mass dimension $d_{\bar{g}_i}$, we define its dimensionless counterpart $g_i = \bar{g}_i\, k^{-d_{\bar{g}_i}}$, i.e., we absorb the dimensionality into an appropriate power of the RG scale $k$.
The fixed point is denoted by $g_{i\, \ast}$, with at least one nonvanishing coupling, i.e., at least one $g_{i\, \ast}\neq 0$, for an interacting fixed point. 

Fixed points act as sources or sinks for the RG flow to the infrared depending on whether RG trajectories flow into a fixed point or emanate from it. Typically, they act as sources of the flow in some directions in the space of couplings while simultaneously acting as sinks in others. Irrelevant directions are associated with sinks. Relevant directions are associated with sources. From the stability matrix,
\begin{align}
	M_{ij}=\dfrac{\del \beta_{g_i} }{\del g_j}  \Big|_{g= g_{\ast}},
\end{align}
we define the critical exponents by $\theta_i = - \lambda_i$, where the $\lambda_i$'s correspond to eigenvalues of the stability matrix. If the real part of a critical exponent is positive, that critical exponent is associated with an eigendirection which corresponds to a \textit{relevant direction}. If it is negative, the critical exponent corresponds to an \textit{irrelevant direction}. 
 
The low-energy value of the couplings associated with a relevant direction\footnote{Strictly speaking, at an interacting fixed point, the eigendirections of a fixed point are often superpositions of couplings.} is a free parameter of the theory and must be determined experimentally. Thus, the theory is predictive if it has a finite number of relevant directions. In gravity, there is compelling evidence that the number of relevant directions is not much higher than three, with most evidence placing the number of relevant directions at three, see \cite{Lauscher:2002sq,Rechenberger:2012pm,Benedetti:2009rx,Benedetti:2009iq,Groh:2011vn,Falls:2013bv,Falls:2014tra,Falls:2017lst,Falls:2018ylp,Kluth:2020bdv,Falls:2020qhj,Knorr:2021slg}. 

In contrast, irrelevant directions give rise to predictions of the low-energy value of the corresponding  couplings. Their values typically depend on the low-energy values of the relevant couplings. Thus one can think of each negative critical exponent as one condition relating couplings to each other. Once the relevant couplings are determined by experiments, the irrelevant couplings are fully determined by solving the flow equation. 

This predictive power of asymptotic safety is one of its remarkable features and could enable observational tests of quantum gravity at energies much below the Planck scale, see Sec.~\ref{sec:gravmatterinterplay} below.

\subsection{The FRG machinery}\label{sec:FRG}

We search for asymptotic safety in gravity-matter systems using the functional Renormalization Group (FRG), for reviews, see \cite{Pawlowski:2005xe,Gies:2006wv, Dupuis:2020fhh}. The FRG allows us to compute the scale-dependence of the couplings of the theory. It is inspired by the Wilsonian approach to the path integral \cite{Wilson:1973jj} which integrates out quantum fluctuations as a function of the RG scale $k$, which is a momentum scale. The central object is the\textit{  flowing action} $\Gamma_k$, in which only quantum fluctuations with momenta larger than $k$ are integrated out. Momenta lower than $k$ are suppressed due to the presence of an \textit{IR regulator} $  \textbf{R}_k(p) $ which is introduced in the generating functional. This regulator function acts like a $k$- and momentum-dependent mass, such that modes with momenta $p^2<k^2$ are made heavy and their fluctuations are suppressed.
The key advantage of this setup is that is gives rise to a functional differential equation, from which one can directly extract the scale dependence of couplings, also beyond the regime of perturbative renormalizability. This equation is
 the \textit{flow equation} \cite{Wetterich:1992yh, Morris:1993qb}, 
\begin{eqnarray}\label{eq:flow}
	k\del_k \Gamma_k = \dfrac{1}{2}\Tr[(\Gamma_k^{(2)}+\textbf{R}_k)^{-1}k \del_k \textbf{R}_k] , \label{flow}
\end{eqnarray}
where the superscript (2)  in $\Gamma_k^{(2)}$ means two functional derivatives with respect to the fields. This equation can be understood as a shorthand notation that summarizes all beta functions of a theory in compact form. To extract individual beta functions, one selects the appropriate field monomials on the right hand side of the equation.

In general, the flowing action $\Gamma_k$, just like the effective action $\Gamma_{k \rightarrow 0}$, contains all field monomials that are compatible with the symmetries, i.e., it contains an infinite tower of operators and corresponding couplings. However, for practical purposes it is typically not necessary to account for this full infinite tower and information up to a certain accuracy can be extracted by truncating $\Gamma_k$. Such a truncation needs to be based on a physical principle. In our case we follow the indications that gravity-matter systems show a near-canonical scaling, i.e., remain near-perturbative \cite{Falls:2013bv,Falls:2014tra,Falls:2017lst,Falls:2018ylp,Eichhorn:2018ydy,Eichhorn:2018nda,Eichhorn:2020sbo} and thus base our truncation on canonical power counting. Nevertheless, choosing a truncation introduces systematic uncertainties. 

Another source of systematic uncertainties is the Euclidean signature, required by the presence of an IR cutoff within the FRG formalism, which is not well-defined in Lorentzian signatures, see \cite{Manrique:2011jc}. If asymptotically safe gravity-matter systems indeed remain near-perturbative, such that an approximately flat background emerges dynamically, a continuation of the Euclidean graviton propagator to Lorentzian signature could be viable \cite{Bonanno:2021squ,Fehre:2021eob}; for analyticity properties of proposed propagators, see \cite{Draper:2020bop,Platania:2020knd}.

\subsection{The interplay of gravity and matter} \label{sec:gravmatterinterplay}
It is not enough to find asymptotic safety in pure gravity because a phenomenologically viable quantum gravity theory must include matter-fields. Even minimally coupled matter fields already change the fixed-point properties in the gravitational sector \cite{Dona:2013qba,Meibohm:2015twa,Dona:2015tnf,Biemans:2017zca,Alkofer:2018fxj,Wetterich:2019zdo,Sen:2021ffc}, with indications that the gravitational fixed point persists under the impact of Standard-Model matter fields and some beyond Standard-Model extensions, including, e.g., an axion \cite{Dona:2013qba}. Moreover, evidence for gravity-matter fixed points is mounting \cite{Eichhorn:2011pc,Dona:2013qba,Dona:2014pla,Meibohm:2015twa,Oda:2015sma,Dona:2015tnf,Wetterich:2016uxm,Eichhorn:2016vvy,Biemans:2017zca,Eichhorn:2017sok,Eichhorn:2017eht,Christiansen:2017cxa,Hamada:2017rvn,Pawlowski:2018ixd,Eichhorn:2018ydy,Alkofer:2018fxj,Eichhorn:2018akn,Eichhorn:2018nda,Wetterich:2019zdo,deBrito:2019epw,deBrito:2019umw,Eichhorn:2020sbo,Burger:2019upn,Kwapisz:2019wrl,Daas:2021abx,Schiffer:2021gwl}; and, more recently, lattice searches for asymptotic safety also included matter fields \cite{Jha:2018xjh,Catterall:2018dns,Dai:2021fqb,Ambjorn:2021fkp,Ambjorn:2021uge}. 

 First, the gravitational parameter space -- spanned by the values of the gravitational couplings in the UV -- is constrained through a mechanism known as \textit{weak-gravity bound} \cite{Eichhorn:2016esv,Christiansen:2017gtg,Eichhorn:2017eht,Eichhorn:2017sok,Eichhorn:2019yzm, deBrito:2021pyi,Eichhorn:2021qet}. This bound prohibits large gravitational strengths because strong gravity fluctuation can trigger new divergences in matter interactions. Thus, gravity in the presence of matter is constrained in it UV properties. 

Second, matter fields could provide a pathway to observational tests of asymptotic safety. The requirement of UV scale symmetry actually imposes constraints on the dynamics of the theory at all scales, even in the IR, where scale symmetry is not realized. This follows, because each irrelevant direction of the fixed point imposes a relation on the couplings of the theory at all scales. If these relations include some of the couplings of the Standard-Model, then predictions of electroweak scale physics become possible based on asymptotic safety at and beyond the Planck scale. Robust indications for this predictive power exist \cite{Harst:2011zx,Eichhorn:2017lry,Eichhorn:2017ylw,Eichhorn:2017als,Pawlowski:2018ixd,Wetterich:2019rsn,
Eichhorn:2020kca,Eichhorn:2020sbo},
and phenomenological scenarios have been worked out on this basis \cite{Shaposhnikov:2009pv,Eichhorn:2018whv,Grabowski:2018fjj,Eichhorn:2019dhg,Reichert:2019car,Alkofer:2020vtb,Kowalska:2020zve,Eichhorn:2020kca,Kowalska:2020gie,Domenech:2020yjf,deBrito:2020dta,Hamada:2020vnf,Eichhorn:2021tsx}. The asymptotically safe landscape is, accordingly, the set of effective field theories for matter fields which is compatible with an asymptotically safe UV completion at high energies.

\subsubsection{The weak-gravity bound}\label{sec:WGB}

Because it will be critical for our results, we explain the weak-gravity bound and the mechanism behind it in more detail. The weak-gravity bound arises, because gravity induces interactions for matter fields \cite{Eichhorn:2011pc,Eichhorn:2012va}. Thus, the corresponding matter couplings cannot be set to zero at a gravitational fixed point. In turn, there is not guarantee for the matter couplings to feature a fixed point under the impact of gravity, at least not if gravity fluctuations are strong. If gravity fluctuations are weak, the matter couplings lie at a shifted Gaussian fixed point, which is a fixed point that would be free in the absence of gravity and that is shifted to finite fixed-point values that depend parametrically on the strength of gravitational fluctuations. However, once gravity fluctuations are strong, the shifted Gaussian fixed point can collide with an interacting fixed point. Upon this collision, both fixed points become complex valued, i.e., they are no longer viable fixed points. This defines a critical value of the gravitational interaction strength, up to which a fixed point in these induced matter interactions exist. Beyond this critical value, there cannot be a consistent gravitational fixed point in gravity-matter systems.  This mechanism is known as \textbf{weak-gravity bound}.

For systems with scalars, 
this bound occurs in shift symmetric interactions \cite{Eichhorn:2016esv,Eichhorn:2017eht,deBrito:2021pyi}, which are necessarily induced by gravity \cite{Eichhorn:2012va,Eichhorn:2013ug,Eichhorn:2017sok,Laporte:2021kyp}. 
In this work, we use the Einstein-Hilbert truncation, in which the dimensionless versions of the Newton coupling $G$ and the cosmological constant $\Lambda$ are the two gravitational couplings. These couplings occur in the combination $G/(1-2\Lambda)^{\#}$ in the matter beta functions (with $\# \geq 1$), and thus gravity becomes strong by either increasing $G$, or by having $\Lambda$ approach $1/2$. Thus, the weak-gravity bound lies at larger $G$, the more negative $\Lambda$ becomes\footnote{Here, $G$ and $\Lambda$ refer to the UV values of these couplings; a negative UV value of $\Lambda$ can be connected to positive $\Lambda$ in the IR and is thus phenomenologically viable.}, cf.~Fig.~\ref{fig:WGBScalars}.

\section{Are there ALPs in the asymptotically safe landscape?}\label{sec:Results}

After introducing ALPs in the first section and the asymptotically safe landscape in the previous section, let us:
\begin{itemize}
\item Minimally couple the ALP action with gravity.
\item Derive the beta functions for this matter-gravity system.
\item Search for fixed points and explore whether there are ALPs in the asymptotically safe landscape.
\end{itemize}

\subsection{ALP-gravity system}\label{sec:settingALP+Grav}

In this section, we minimally couple the ALP action to gravity and choose a truncation of the effective action of the form
\begin{align}
	\Gamma_k= \Gamma_{k,\, \rm grav} + \Gamma_{k,\, \rm matter} + \Gamma_{k,\text{gf}} \,.
\end{align}
For the gravitational dynamics, we write a Euclidean Einstein-Hilbert term 
\begin{align}
	\Gamma_{k,\, \rm grav}=\dfrac{1}{16\pi G\, k^{-2}}\int\sqrt{\det g }\left(2\Lambda\, k^2-R\right),
\end{align}
which we have written in terms of the dimensionless Newton coupling $G$ and the dimensionless cosmological constant $\Lambda$. For the ALP action in the presence of a dynamical metric, we write
\begin{align}
\Gamma_{k,\, \rm matter}= \int d^4 x\sqrt{\det g}\,&\left( \dfrac{Z_\phi}{2}g^{\mu\nu}\del_\mu \phi\del_\nu\phi+ \dfrac{{Z_\phi\,m^2 k^2}}{2} \phi^2 \right. \nn  \\ & \left. + \dfrac{Z_A}{4} g^{\mu\alpha}g^{\nu\beta}F_{\mu\nu}F_{\alpha\beta}  + \dfrac{i Z_\phi^{1/2} Z_A\, g
 k^{-1}}{8} \dfrac{\epsilon^{\mu\nu\alpha\beta}}{\sqrt{\det g}} \,\phi \, F_{\mu\nu}F_{\alpha\beta} \right).
\label{eq:alpgravity}
\end{align}
This truncation is motivated by phenomenological considerations at low energy:  the parameter space for ALPs that is investigated experimentally is spanned by the dimensionful mass $\bar{m} = m\, k$ and the dimensionful ALP-photon coupling $\bar{g}= g \, k^{-1}$, which we therefore include in our truncation.
Additionally, we include kinetic terms for the scalar and the photon, as well as the two lowest-dimensional terms for gravity. All ALP terms satisfy parity symmetry.

We supplement the gravitational and photon actions by gauge fixing terms. For the gravitational part, we use a background gauge fixing and split the metric into a background, $\bar{g}_{\mu\nu}$ (which we choose to be flat, $\bar{g}_{\mu\nu} = \delta_{\mu\nu}$) and a fluctuation field, $h_{\mu\nu}$,
\begin{align}
	g_{\mu\nu}=\bar{g}_{\mu\nu} + \left(32 \pi\, G k^{-2}\, Z_h \right)^{1/2}  h_{\mu\nu},
\end{align}
where $Z_h$ is a wave-function renormalization for the fluctuation field.
The $U(1)$ and gravitational gauge-fixing terms are given by
\begin{align}
	\Gamma_{k,\text{gf}}=\dfrac{1}{2\alpha_A}\int d^4 x \sqrt{\det \bar{g}} \,(\bar{g}^{\mu\nu} \, \bar{D}_\mu A_\nu)^2 + \dfrac{1}{\alpha}\int d^4 x \sqrt{\det \bar{g}} \,\bar{g}_{\mu\nu}\mathcal{F}^\mu\mathcal{F}^\nu,
\end{align}
where $\bar{D}^\alpha $ is the covariant derivative defined with respect to the background metric $\bar{g}_{\mu\nu}$, and
\begin{align}
	\mathcal{F}^\mu = \left( \bar{g}^{\mu\nu} \bar{D}^\alpha - \dfrac{1+\beta}{4} \bar{g}^{\nu\alpha} \bar{D}^\mu \right)h_{\nu\alpha}.
\end{align}
We choose the Landau-deWitt gauge $\alpha=\beta=0$; and Landau gauge $\alpha_{A}=0$ for the $U(1)$ sector.

For the gravitational dynamics, our choice of truncation neglects canonically marginal and irrelevant higher-derivative operators compatible with diffeomorphism symmetry. It has been shown in the literature, that the existence of a non-trivial fixed point in the purely gravitational sector, with a finite number of relevant directions is robust under such extensions of the truncation \cite{Lauscher:2002sq,Codello:2008vh,Benedetti:2009rx,Benedetti:2009iq,Groh:2011vn,Rechenberger:2012pm,Falls:2013bv,Falls:2014tra,Gies:2016con,Falls:2017lst,Falls:2018ylp,Kluth:2020bdv,Falls:2020qhj,Knorr:2021slg}. Since these operators contribute to the beta functions for the matter couplings, such operators might affect the effect of gravity on ALPs. We comment on this point in our conclusions/outlook. 

Our choice of truncation also neglects non-minimal couplings. First, we neglect nonminimal couplings that satisfy shift-symmetry, such as 
$R^{\mu\nu}\,\del_\mu\phi\del_\nu\phi$, which have been investigated in \cite{Eichhorn:2017sok}, and  $R \,g^{\mu\nu}\partial_{\mu}\phi \partial_{\nu}\phi$, see also \cite{Laporte:2021kyp}. These couplings are expected to be nonzero, but their impact on the matter couplings in our truncation is expected to be small and through the anomalous dimension, only. Second, we neglect the canonically marginal nonminimal coupling $R\phi^2$ which breaks shift symmetry. As long as the ALP-mass has a vanishing fixed-point value, it is consistent to set the nonminimal coupling to zero. At large values of the ALP mass, our approximation is expected to be less quantitatively reliable.
However, our main interest in this work is in investigating how gravity deforms the Gaussian fixed point of the matter couplings, so we will focus on small values for the couplings $g$ and $m$. 

For the matter dynamics, our choice of truncation neglects any other self-interaction invariant under parity and $ U(1) $  than the one characterized by the ALP-photon dimensionless coupling $g$. In our truncation, the canonically marginal operator $\lambda \phi^4$ can only be generated in the presence of gravity, and this contribution is suppressed since it depends on $m^4$. The same argument holds for the canonically irrelevant operators in the ALP potential. 

To summarize, we do not expect that the inclusion of canonically irrelevant operators in our truncation will change the fate of ALPs in the asymptotically safe landscape, at least in the region of small couplings. Therefore, our results cover more general ALP models than the one for which the potential is given by $ V(\phi) =  \frac{\bar{m}^2}{2} \phi^2$.

Finally, the regulator term is diagonal in field space and its components are given by
\begin{align}
	\textbf{R}_k(p^2)=
	\left( \Gamma_k^{(2)}(p^2) - \Gamma_k^{(2)}(0)\right)_{h,\phi,A=0}\, r_k(p^2/k^2).
\end{align}
In this work, we chose the Litim regulator \cite{Litim:2001up}, given by $ r_k(y)=\left(\frac{1}{y}-1\right)\theta(1-y)$. Such a regulator respects all the symmetries under which the interactions we consider are invariant. 

We evaluate the scale dependence of the matter coupling through a vertex expansion of the flow equation \eqref{eq:flow}. 
To derive the vertices that enter the flow equation, we only need to expand the effective action to second order in $h_{\mu\nu}$. In this way, we obtain the graviton propagator and the vertices $h\phi\phi $, $h^2\phi\phi $, $hAA $ and $h^2AA $.

In Eq.~\eqref{eq:alpgravity}, we highlight that the presence of the Levi-Civita symbol cancels the determinant of the metric.\footnote{Remember that the antisymmetric Levi-Civita symbol is a tensor density. Therefore, in the presence of a dynamical metric, it depends on the determinant of the metric, such that the combination $\sqrt{\det g}\,F\tilde{F}$ is a topological term.} Consequently, the only interaction vertex that contains both $\phi$ and $A$ is $\phi AA$, even when gravity is present. 

\subsubsection{Beta functions}\label{sec:betafunctions}

In this section, we discuss the beta functions associated with the truncation from the previous section. By using the FRG flow equation introduced in Sec.~\ref{sec:FRG}, we can extract equations for the anomalous dimensions and beta functions of the matter couplings by using suitable projections. We use self-written Mathematica codes based on the packages \textit{xAct}  \cite{Brizuela:2008ra,Martin-Garcia:2007bqa,MartinGarcia:2008qz}, \textit{FormTracer} \cite{Cyrol:2016zqb}, \textit{DoFun} \cite{Huber:2019dkb} and \textit{Package-X} \cite{Patel:2016fam} to derive the beta functions.

To compute the scale dependence, we make an assumption about the anomalous dimensions. The assumption relates to the typical structure of FRG flow equations, which come with $\left(\#- \eta_i\right)$ factors in the numerators, where $\#>1$ and the anomalous dimension $\eta_i$ arises because the regulator is chosen to be proportional to $Z_i$, see App.~\ref{sec:betaApp} for more details. Here,
we neglect the anomalous dimensions that arise in this way. First, that means that we set the graviton anomalous dimension $\eta_h=0$ everywhere. Second, that means that we set $\eta_{\phi} = 0 = \eta_A$, except for their appearance in the canonical terms of the beta functions. This approximation simplifies the structure of the beta functions, which only depend polynomially on the coupling $g$. It holds, as long as the anomalous dimensions do not become too large -- a condition that we can check on any fixed points we find in the beta functions.

 The RG-flow for the ALP-mass and for the ALP-photon coupling is described by the following beta functions
\begin{align}
	\beta_{m^2} & = -2m^2 + \frac{m^2 g^2}{16\pi^2} 
	+ \frac{5 \, m^2\,G}{2\pi (1-2\Lambda)^2} + \frac{m^2\,G}{3\pi (1-4\Lambda/3)^2} 
	+ \frac{m^2 \,G}{18\pi (1+m^2)(1-4\Lambda/3)^2}\nonumber\\
	& 
	+ \frac{(1 - 24\,m^2)\,m^2\,G}{18\pi (1+m^2)^2(1-4\Lambda/3)}
	- \frac{4\,m^6 \,G}{3\pi (1+m^2)^2(1-4\Lambda/3)^2} \,, \label{eq:betam2gravity} \\
	 \beta_{g^2} & = 2g^2 + \frac{g^4}{16\pi^2} + \frac{g^4}{24\pi^2(1+m^2)} + \frac{g^4}{24\pi^2(1+m^2)^2}  
	-\frac{20 \,g^2\, G}{9\pi(1-2\Lambda)} \nonumber \\
	&+ \frac{10\,g^2\,G}{9\pi(1-2\Lambda)^2}
	+ \frac{(1+24m^2)\,g^2 \,G}{18\pi (1+m^2) (1-4\Lambda/3)^2}
	+ \frac{g^2 \,G}{18\pi (1+m^2)^2 (1-4\Lambda/3)}\nonumber \\
	&-\frac{4\,m^4\,g^2 \,G}{3\pi (1+m^2)^2 (1-4\Lambda/3)^2} \label{eq:betag2gravity}\,.
\end{align}
For additional details on the structure of the beta functions $\beta_{m^2}$ and $\beta_{g^2}$, see App.~\ref{sec:betaApp}.

\subsection{ Fixed points in the matter sector under the impact of gravity} \label{subsec:fixedpoints}

Can gravitational interactions induce fixed points phenomenologically compatible with ALPs? In this section, we aim to offer the first answer to this question within asymptotically safe quantum gravity. We treat the gravitational couplings as free parameters, such that we arrive at a complete understanding of the gravitational parameter space. From \cite{Dona:2013qba}, it is known that there is a gravitational fixed point, also if the ALP and the photon are coupled to gravity minimally. Thus, we denote the gravitational fixed-point values by $G_{\ast}$ and $\Lambda_{\ast}$, but keep their values free.

 From Eqs.~\eqref{eq:betam2gravity} and \eqref{eq:betag2gravity}, we observe that gravity has a screening effect on the mass and an antiscreening effect on the ALP-photon coupling at their respective free fixed points: The canonical critical exponents, $2$ for the $m^2$ and $-2$ for $g^2$, are each counteracted by a gravitational term. For the mass, these last three terms in the first line of Eq.~\eqref{eq:betam2gravity} which are $\sim m^2\, G_{\ast}$ are positive for all $\Lambda \in (-\infty,0.5)$, counteracting the canonical term. Similarly, for the ALP-photon coupling, the terms $\sim g^2\ G_{\ast}$ in Eq.~\eqref{eq:betag2gravity} are negative for all $\Lambda \in (-\infty,0.24)$. This is a crucial result, because it implies that gravitational fluctuations have the correct sign to enable a UV completion of the ALP-photon interaction: the antiscreening effect of gravity, if it is large enough, can render the ALP-photon coupling asymptotically free. 

\subsubsection{Free fixed point} \label{subsec:freefixedpoint}

From Eqs.~\eqref{eq:betam2gravity} and \eqref{eq:betag2gravity}, it is straightforward to see that the free fixed-point $(g^2_*,m^2_*)=(0,0)$ is a solution of the fixed-point equations $\beta_{m^2}=0$ and $\beta_{g^2}=0$. This is in line with expectations based on symmetry considerations:
 When we couple matter to gravity, free fixed points can be shifted to become interacting fixed-points in the presence of gravity. Such a shift occurs for interactions which are invariant under all symmetries of the kinetic operator \cite{Eichhorn:2017eht}.  In our case, the ALP-photon coupling breaks the $\mathbb{Z}_2$ symmetry of the ALP kinetic term and the mass breaks the shift symmetry of the ALP kinetic term. Thus, the existence of a fixed point at $(g^2_*,m^2_*)=(0,0)$ is expected.

\begin{figure}[!t]
	\centering
	\hspace*{-.5cm}\includegraphics[scale=.95]{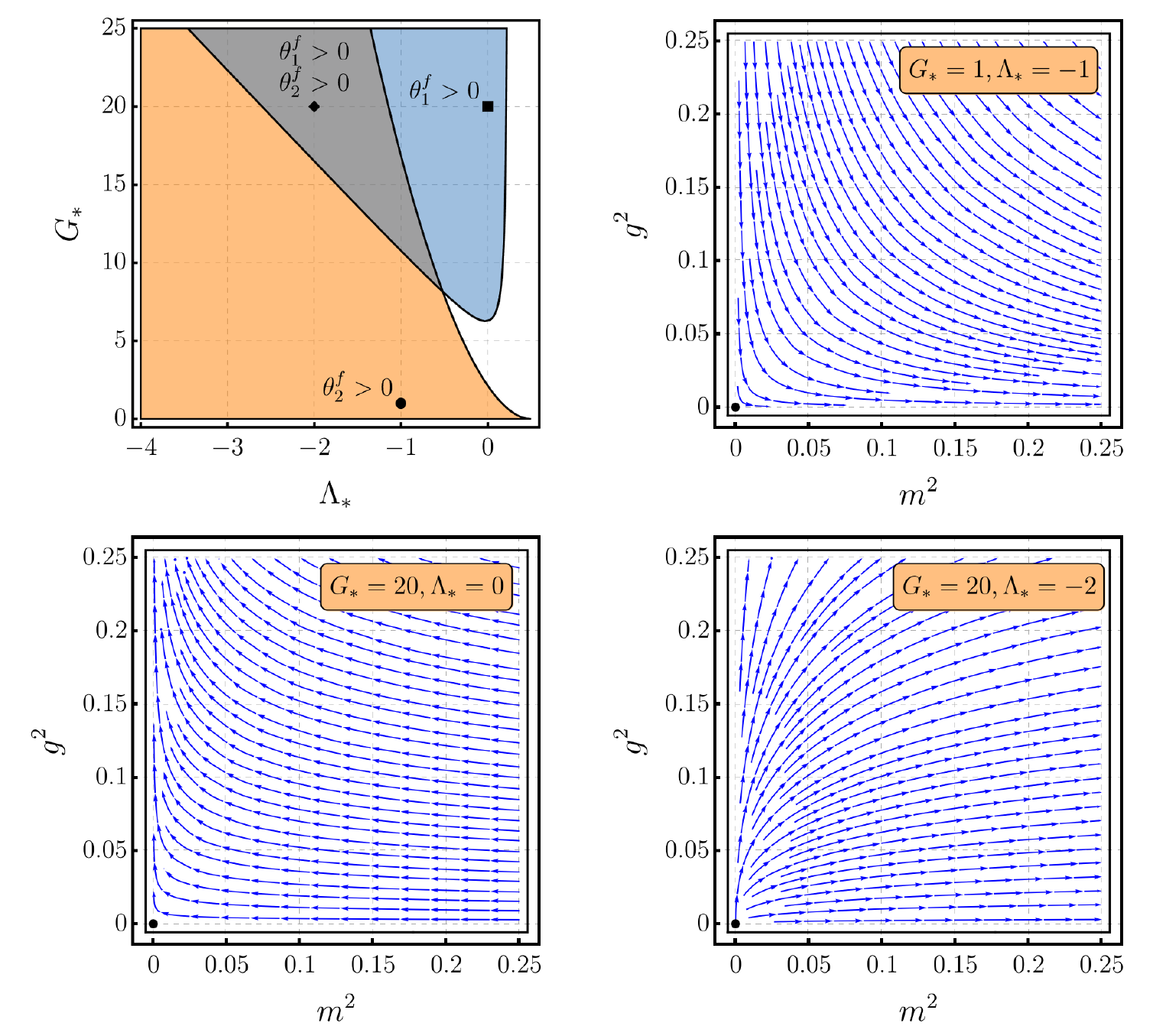}
	\caption{\label{fig:FreeFPcouplings}  In the upper left panel, we show the sign of the critical exponents associated with the free fixed point in the presence of gravity. In blue, we highlight the parameter space where $\theta_1^f>0$, in orange, the region where $\theta_2^f>0$. In the other three panels, we show the RG flow towards the IR in the vicinity of the free fixed point for the three different regions in the upper left panel; with values for $G_{\ast}, \Lambda_{\ast}$ indicated by a dot (corresponding to the upper right panel), a square (corresponding to the lower left panel) and a diamond (corresponding to the lower right panel). }
\end{figure}

Whether or not this fixed point is consistent with an interacting, massive ALP at low energies, depends on the critical exponents at this fixed point.
The critical exponents associated with the free fixed-point are 
\begin{eqnarray}
	\theta_1^f &=& -2+\frac{G_* \left(-616 \Lambda_* ^3+1060 \Lambda_* ^2-558 \Lambda_* +81\right)}{9 \pi  \left(8 \Lambda_* ^2-10 \Lambda_* +3\right)^2}, \label{eq:theta1gfp}\\
	\theta_2^f &=& 2+\frac{G_* \left(16 \Lambda_* ^3-352 \Lambda_* ^2+460 \Lambda_* -159\right)}{6 \pi  \left(8 \Lambda_* ^2-10 \Lambda_* +3\right)^2},\label{eq:theta2gfp}
\end{eqnarray}
where $\theta_1^f$ is associated with the ALP-photon coupling, and $\theta_2^f$ is associated with  $m^2$. The ALP-photon coupling is canonically irrelevant and
for small values of $G_{\ast}$, gravity is not strong enough to change the sign of  $\theta_1^f$. The second critical exponent $\theta_2^f$ is positive, dominated by the canonical dimension of the mass, at small values of $G_{\ast}$.
Thus, the low-energy phenomenology resulting from this fixed point at weak gravity (small $G_{\ast}$ is a massive ALP with a vanishing ALP-photon coupling. This can also be seen from the upper right panel in Fig.~\ref{fig:FreeFPcouplings}, where the RG flow from the free fixed point towards the IR is towards growing $m^2$, but remains at $g^2=0$.

However, for $\Lambda_{\ast}\lesssim 0.24$, the gravitational contribution is positive. Thus, when $G_{\ast}$ is sufficiently big, the gravitational contribution overwhelms the canonical scaling, and $g$ becomes relevant.  Then, both $g^2$ and $m^2$ can correspond to relevant directions, such that the RG flow away from the free fixed point can reach any combination of values of $\bar{g}$, $\bar{m}^2$ in the IR. This scenario is realized in the overlap of the orange and the blue region in the upper left panel in Fig.~\ref{fig:FreeFPcouplings}.

For even larger $G_{\ast}$, the negative gravitational contribution in Eq.~\eqref{eq:theta2gfp} overwhelms the canonical scaling and the mass becomes irrelevant, cf.~lower left panel in Fig.~\ref{fig:FreeFPcouplings}. The resulting IR phenomenology features an ALP-photon coupling for a massless ALP. Since ALPs are motivated to be dark-matter candidates, they cannot be massless; otherwise, they would produce relativistic (hot) dark-matter, which is in tension with structure formation. \\

\begin{figure}[!t]
	\centering
	\hspace*{-1.cm}
	\includegraphics[scale=.975]{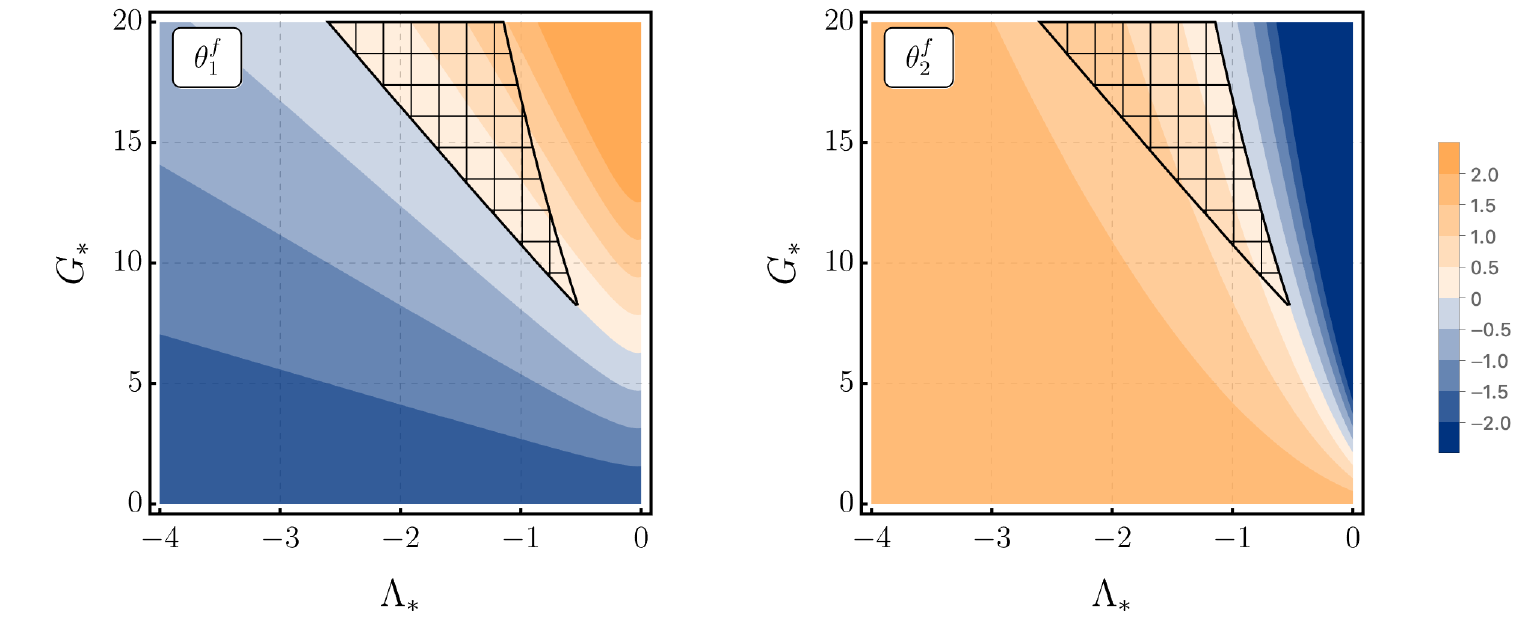}
	\caption{\label{fig:FreeFPcritexps}  We show the 
	scaling exponents for the square of ALP-photon coupling and for the ALP mass squared at the free fixed point $(m_*^2,g_*^2)=(0,0)$. In the meshed region both $m^2$ and $g^2$ are relevant couplings.}
\end{figure}

To check the self-consistency of the truncation underlying our results, we check the values of the critical exponents. They should not show very significant deviation from their canonical values because our truncation is based on canonical power counting. According to Fig.~\ref{fig:FreeFPcritexps}, in the region we are interested in, the critical exponents for $m^2$ and  $g^2$ (which have canonical scaling dimension $+2$ and $-2$, respectively) deviate from the canonical scaling by no more than 4. Here, it becomes important that we are looking at squares of couplings here; for the couplings $g$ and $m$, the deviation is no more than 2, which says that strongly irrelevant couplings are not expected to become relevant. Thus, we consider our truncation to be sufficient for this first study of the gravity-ALP interplay.

 Nevertheless, the phenomenologically interesting region -- where nonzero ALP-photon couplings and ALPs masses can be achieved, is in a regime where gravity is already interacting relatively strongly. This can be seen from the fact that the gravitational contribution to the scaling dimension is large enough to overwhelm the canonical scaling. This region of strongly-interacting quantum gravity has been shown to be in tension with the existence of certain matter fields, because of the weak-gravity bound, discussed in Sec.~\ref{sec:WGB}.

The result from \cite{deBrito:2021pyi} for the weak-gravity bound holds for pseudo-scalar fields because the interactions that are involved are the same for scalars and pseudoscalars. Therefore, not all of the parameter space we considered in Fig.~\ref{fig:FreeFPcouplings} is in fact available: while the free fixed point for  $g^2$ and $m^2$ exists for all $G_{\ast}$ and $\Lambda_{\ast}$, gravity triggers divergences in shift-symmetric ALP-interactions of the form $(\partial_{\mu} \phi \partial^{\mu}\phi)^2$, if the strength of gravity exceeds the weak-gravity bound. 

\begin{figure}[!t]
	\centering
	\includegraphics[width=0.5\linewidth]{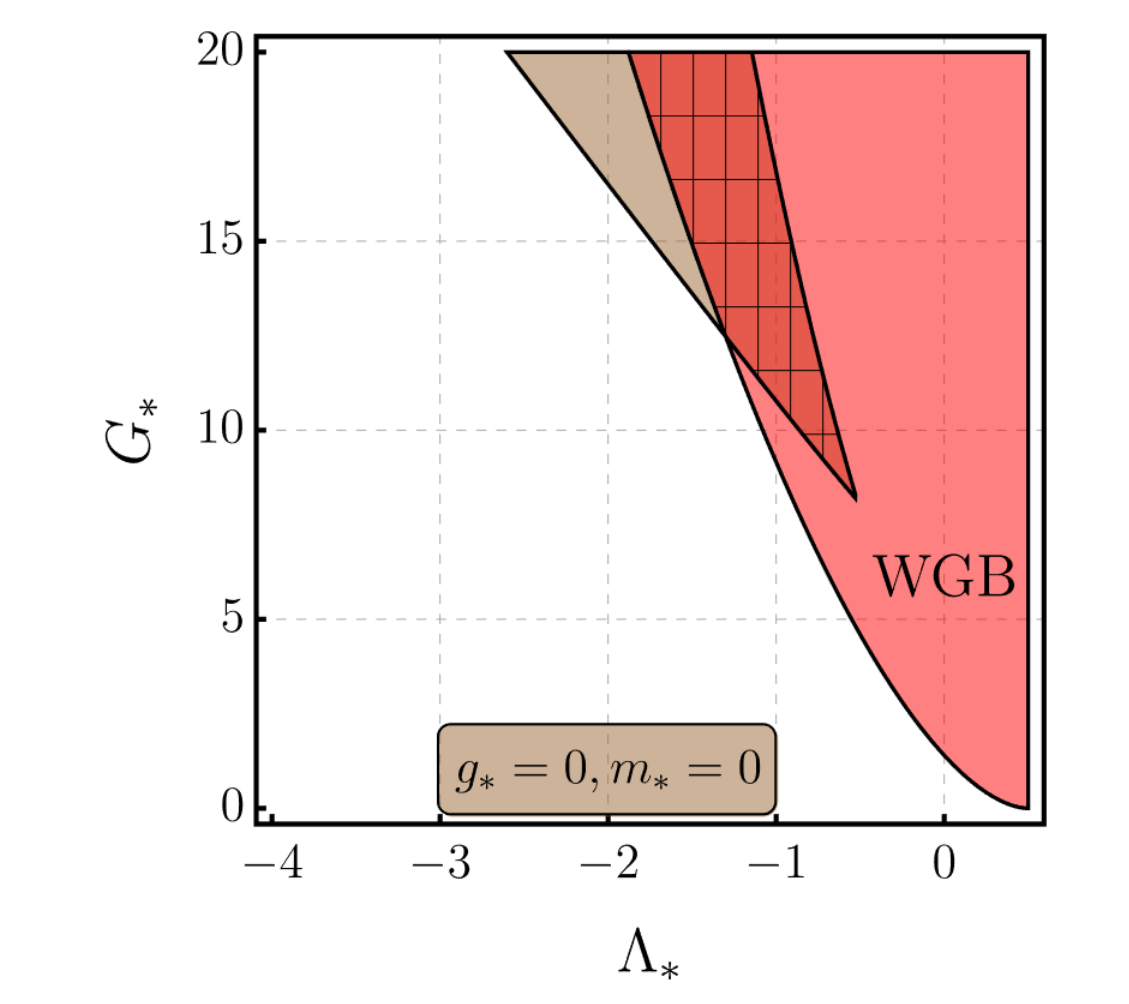}
	\caption{\label{fig:WGBScalars} In red, we show the region which is forbidden by the WGB and in brown the region where the free fixed point at $ g_*=0,m_* = 0 $ has two relevant directions. Only the brown region without the meshing style is allowed by the WGB.} 
\end{figure}

For the (pseudo-)scalar-gravity truncation from \cite{deBrito:2021pyi}, the weak-gravity bound is given by the expression
\begin{eqnarray}
	G_{\rm c}(\Lambda_{\ast})= \frac{8\pi}{ \frac{20/3}{(1-2 \Lambda_{\ast} )^2} - \frac{8/9}{(1-4 \Lambda_{\ast}/3)} -\frac{5/9}{(1-4 \Lambda_{\ast}/3)^2} + \sqrt{\frac{160}{(1-2 \Lambda_{\ast})^3}+\frac{1}{(1-4 \Lambda_{\ast}/3)^2}+\frac{3}{(1-4 \Lambda_{\ast}/3)^3}}}.
	\quad\label{eq:GcritLambda}
\end{eqnarray}
In Fig.~\ref{fig:WGBScalars}, we show the weak-gravity bound, as well as the region in which the free fixed point has two relevant directions. There is an overlap between both regions, such that the region in which an ALP with nonzero mass and nonzero coupling to photons could exist, is reduced very significantly and potentially disappears completely: within our truncation, a viable region remains; however, the boundaries in the space of gravitational couplings are subject to systematic uncertainties (related to the choice of truncation). Therefore, within the systematic uncertainties, it is unclear whether or not a small region of the gravitational parameter space remains. 

Our final observation is that besides constraints that arise from requiring a nonzero ALP-photon coupling as in Fig.~\ref{fig:WGBScalars}, there are constraints that arise from requiring non-zero Standard-Model matter couplings, most importantly Yukawa couplings \cite{Oda:2015sma,Eichhorn:2016esv,Eichhorn:2017ylw,Eichhorn:2020sbo}. The corresponding region lies at $\Lambda_{\ast}<-3.3$, such that the combination of ALP-constraints with Yukawa constraints results in a region at rather large $G_{\ast}$.
We point out that in previous work on gravity-matter systems, the gravitational fixed-point values $G_{\ast}, \Lambda_{\ast}$ have been determined -- albeit with systematical uncertainties -- to lie at significantly lower values of $G_{\ast}$ for negative $\Lambda_{\ast}$, than would be required to reach this phenomenologically interesting region. In summary, this suggests that the region with a finite ALP coupling is in tension with various other requirements for the gravitational fixed point.

\subsubsection{Interacting, massless fixed point}

From Eq.~\eqref{eq:betam2gravity}, we can see that $m^2_\ast=0$ is a solution for any value of $g^2_*$. Given that $m^2$ is canonically relevant, it is likely to stay canonically relevant at small enough values of gravitational couplings. Thus, we search for a fixed point at $g_{\ast}\neq 0, \, m_{\ast}=0$ next. The existence of interacting fixed points that are phenomenologically viable  would be an effect of gravitational quantum corrections since they do not exist for the pure-matter case, see App.~\ref{sec:purematter}. We set $m^2=0$ in Eq.~$\eqref{eq:betag2gravity}$ and obtain the fixed-point value
\begin{align}
	g^2_*=\frac{16}{21}\pi \left(-18 \pi  +\frac{G \left(-616 \Lambda ^3+1060 \Lambda ^2-558 \Lambda +81\right)}{\left(8 \Lambda ^2-10 \Lambda +3\right)^2} \right) \label{eq:g2fpm=0}.
\end{align}
This must be a positive number. Thus, Eq.~\eqref{eq:g2fpm=0} defines a region in the parameter space, where this candidate for an interacting fixed point can exist. By comparison with Eq.~\eqref{eq:theta1gfp}, we see that demanding $g_*^2>0$ is equivalent to demanding $\theta_1^f>0$. Therefore, the blue region in the upper-left panel of Fig.~\ref{fig:FreeFPcouplings} coincides with the region where a candidate for interacting fixed-point can exist.

The mechanism behind this coincidence is well-known for matter-gravity systems, where interacting fixed points for matter couplings exist as soon as the matter couplings become relevant at the free fixed point \cite{Eichhorn:2017ylw, Eichhorn:2017lry, Eichhorn:2018whv}. We can see this mechanism from a schematic form of the beta function,
\begin{align}
	\beta_{g^2}\sim g^2 ( - \theta_1^f + B g^2), 
\end{align}
where $B$ is positive. One can see that the first term corresponds to the critical exponent (associated with $g_*^2=0$) by noting that $\theta_1^f = - \frac{\partial \beta_{g^2}}{\partial g^2}\vert_{g^2=0}$. 
 The existence of an interacting fixed-point requires that $\theta_f$ to be positive, since $ g_* = \sqrt{\theta_1^f/B} $. This is exactly the requirement that $g$ is relevant at the free fixed point. Moreover, the value of the critical exponent $\theta_1^f$ is minus the value of the critical exponent at the interacting fixed point\footnote{At interacting fixed points, critical exponents are typically associated to superpositions of couplings. In the present case, since $m_{\ast}=0$, the stability matrix is diagonal also for $g_{\ast}\neq 0$ and thus the critical exponents are associated to $g$ and $m$, not superpositions of the two.}. We call this critical exponent $\theta_1^i$. Hence, if $g$ is relevant at the free fixed point, it is irrelevant at the interacting one. Thus, the theory is UV completed while predicting a non-vanishing IR value of $g^2$. The same mechanism has been discovered for the Abelian gauge coupling \cite{Harst:2011zx, Eichhorn:2017lry} and Yukawa couplings \cite{Eichhorn:2017ylw,Eichhorn:2018whv}.

\begin{figure}[!t]
	\centering
	\hspace*{-.5cm}\includegraphics[height=7.0cm]{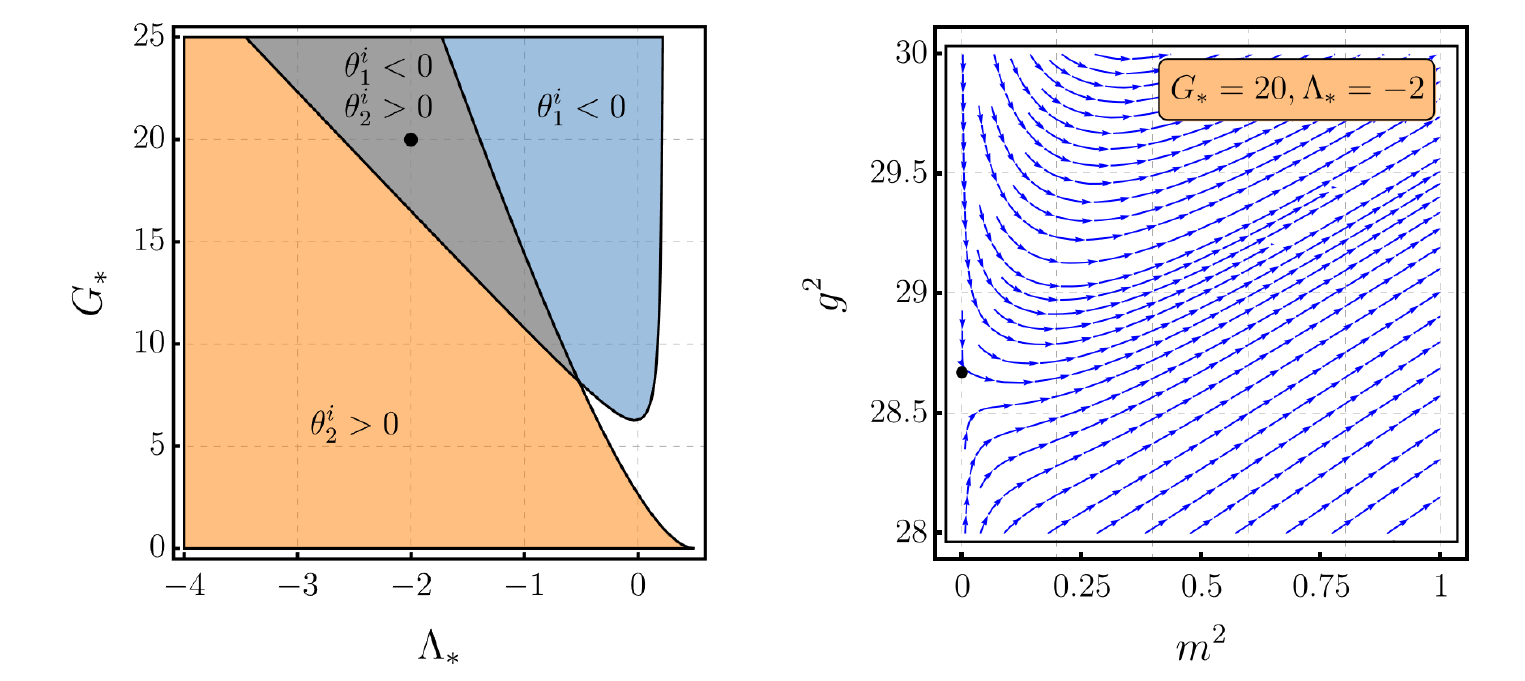}
	\caption{\label{fig:IFP} On the left, in blue, we highlight the parameter space where there are solutions for interacting-fixed points in the presence of gravity. At the same region, $\theta_1^i<0$. In orange, the region where $\theta_2^i>0$. On the right, we show a stream plot around the interacting fixed-point, for $ G_*=20,\Lambda_*=-2$. The black dot indicates the location of the interacting fixed-point. The arrows and lines represent the flow from UV to IR, showing that the $m^2$ direction is relevant and $g^2$ is irrelevant. }
\end{figure}

The second critical exponent, $\theta_2^i$, can be either negative or positive, as shown in the left-panel of Fig.~\ref{fig:IFP}. Again, we are only interested in the region where the mass is associated with a relevant direction; otherwise, it would flow towards zero in the IR, which is phenomenologically unfavored. The critical exponents are
\begin{eqnarray}
	\theta_1^i &=& 2+\frac{G \left(616 \Lambda ^3-1060 \Lambda ^2+558 \Lambda -81\right)}{9 \pi  \left(8 \Lambda ^2-10 \Lambda +3\right)^2}, \label{eq:theta1fpm=0}\\
	\theta_2^i &=& 2+\dfrac{6}{7}+ \frac{G \left(1344 \Lambda ^3-4584 \Lambda ^2+4336 \Lambda -1275\right)}{42 \pi  \left(8 \Lambda ^2-10 \Lambda +3\right)^2}.\label{eq:thetasintFPzeromass}
\end{eqnarray}
The left panel of Fig.~\ref{fig:IFP} shows that there is a region where $\theta_1^i<0$ and $\theta_2^i>0$. This is the region we are interested in: the $g^2$ direction is irrelevant and linked to a calculable, finite value of the ALP-photon coupling in the IR, and $m^2$ is relevant, as shown in the right panel of Fig.~\ref{fig:IFP}.

\begin{figure}[!t]
\centering
\includegraphics[width=0.5\linewidth]{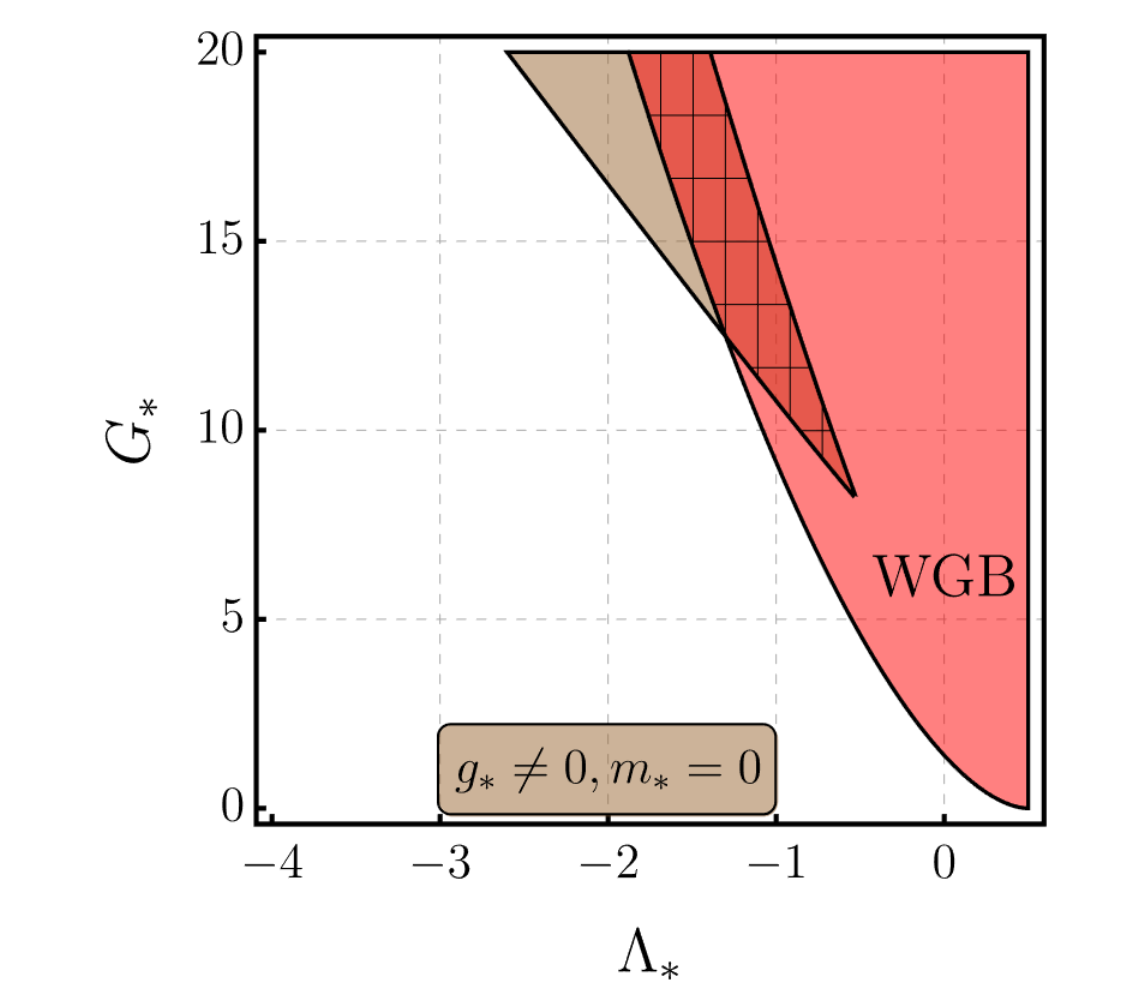}
\caption{\label{fig:WGBint} We show in brown the region where the interacting, massless fixed point exists at the same time the mass corresponds to a relevant direction. The region excluded by the weak-gravity bound is shown in red. In comparison to Fig.~\ref{fig:WGBScalars}, the lower/left boundary of the brown region is the same, but the upper/right is shifted, because the critical exponent associated with the mass, Eq.~\eqref{eq:thetasintFPzeromass}, receives a contribution from $g_{\ast}\neq 0$ which is absent at the fixed point in Fig.~\ref{fig:WGBScalars}.}
\end{figure}

However, because the region in which the fixed point exists is the same in which the photon-ALP coupling is relevant at the free fixed point, the same considerations regarding the weak-gravity bound, compatibility with other phenomenological constraints and location of the gravitational fixed point apply. Thus it is unlikely that viable parameter space remains for this fixed point, cf.~Fig.~\ref{fig:WGBint}.

\subsubsection{Search for other fixed points}

When we set $g=0$ to search for massive, non-interacting fixed point, our setting reduces to that of a scalar field coupled to gravity, which has been investigated thoroughly in \cite{Narain:2009fy, Percacci:2015wwa,Labus:2015ska,Eichhorn:2017als,Pawlowski:2018ixd,Eichhorn:2020sbo}. No indications have been found for a robust fixed point at nonvanishing potential.\\

Further, we find no indications for fully interacting fixed points $(g_*\neq 0, m_*\neq 0 )$ that are compatible with the weak-gravity regime.

\subsection{Steps towards more realistic scenarios}

A caveat of our results is their basis on a toy model, i.e., we focused on the ALP-photon interaction while neglecting other Standard-Model interactions. A more realistic investigation should embed the ALP within the Standard Model. We discuss how to extend our analysis in the future to test how robust our conclusions are.
\\

It is central to keep in mind that the fixed point at $g_{\ast}=0$ and vanishing ALP-potential will persist, based on a symmetry argument: due to shift symmetry, which is not broken by quantum gravity, the fixed point is guaranteed to exist. Therefore, we discuss whether interactions in the Standard Model, which were not part of our toy-model-study, can change the critical exponent of $g$ at this fixed point and make it relevant, which would change important conclusions of our study.

In a more realistic scenario, where the ALP is embedded within the Standard Model, there are charged fermions which generate a contribution to the anomalous dimension of the $U(1)_Y$ gauge field. Thus, the beta function for the ALP-coupling (Eq.~\ref{eq:betag2gravity}) receives the additional contribution
\begin{align}
	\Delta \beta_{g^2}|_{U(1)_Y} = \# \, g_Y^2 \, g^2 \,,
\end{align}
where $g_Y$ denotes the hypercharge gauge coupling and $\#$ represents a positive number. According to \cite{Harst:2011zx, Eichhorn:2017lry}, there are two possibilities for UV completion in the hypercharge sector based on the gravity-contribution to $\beta_{g_Y}$ \cite{Daum:2009dn, Harst:2011zx, Folkerts:2011jz,Christiansen:2017gtg,Eichhorn:2017lry,Christiansen:2017cxa,Eichhorn:2019yzm,deBrito:2019umw}: 

i) $g_{Y}$ is asymptotically free. In this case, there is no additional contribution to $\beta_g$ and our conclusions remain unchanged.

ii) $g_Y$ becomes asymptotically safe at a gravity-induced interacting fixed-point with $g_{Y,*} \sim \sqrt{G_*}$. 
Because $\Delta \beta_{g^2}|_{U(1)_Y}$ is strictly positive for $g_{Y,*} \neq 0$, it pushes the critical exponent $\theta_1^f$ towards negative values. 
Therefore, according to the discussion in Sec.~\ref{subsec:fixedpoints}, this setting further disfavors the possibility of ALPs within the asymptotically safe landscape.
\\

Other directions in which we can extend our analysis include dimension-5 operators involving Standard Model fermions interacting with ALPs as discussed in \cite{Bonilla:2021ufe}. The analysis including this class of interactions goes beyond the scope of this work, and therefore we will keep it for future investigation. 

\section{Conclusions and outlook}\label{sec:conclusions}
We have for the first time explored the impact of asymptotically safe quantum gravity on the ALP-photon coupling. At a first glance, this impact would seem negligible, given the significant gap between the Planck energy and energies at which experimental searches for ALPs are conducted. However, as we have argued, quantum gravity need not to be dynamically important at experimental energy scales in order to be of relevance for constraints on the ALP-photon coupling: A consistent embedding into a fundamental theory including gravity restricts the allowed low-energy values of the ALP-photon coupling. We have shown this to be the case for a toy model of the Standard-Model, where we have neglected all sectors except the Abelian one. Thus, there is no electroweak symmetry breaking and the Abelian gauge group at low and high energies is the same. \\

In more detail, we have found that there are two distinct regimes in the quantum gravitational parameter space: 

In the weakly coupled gravity regime, gravity fluctuations are not strong enough to overcome the canonical scaling of the ALP-photon coupling, which remains irrelevant at the free fixed point. This translates into a prediction of a vanishing low-energy value of the coupling.

In the strongly coupled gravity regime, gravity fluctuations are strong enough to overcome the canonical scaling of the ALP-photon coupling, which becomes relevant at the free fixed point, i.e., it becomes asymptotically free. At the same time, this change in scaling dimensionality induces an interacting fixed point for the ALP-photon coupling, at which its low-energy value is predictable.\\

Together, the structure of these two fixed points would imply that the low-energy value of the ALP-photon coupling is bounded from above, similar to what happens for the Abelian gauge coupling \cite{Christiansen:2017gtg,Eichhorn:2019yzm} or Yukawa couplings \cite{Oda:2015sma,Eichhorn:2016esv,Eichhorn:2017ylw, CMS:2019art,Agaras:2020zvy} in asymptotic safety. However, as a key difference to these other couplings, gravity fluctuations need to be strong in order to induce such a fixed-point structure. In turn, the strong-gravity regime is constrained by the weak-gravity bound, discovered in previous work \cite{Eichhorn:2016esv, Eichhorn:2017eht,Christiansen:2017gtg,Eichhorn:2019yzm,deBrito:2021pyi,Eichhorn:2021qet}, which prohibits asymptotically safe gravity-matter models from entering the strong gravity regime. The reason is that in this regime, gravity triggers novel divergences in higher-order matter interactions, and thus no viable fixed point for the gravity-matter model can exist.

We find that a slice of parameter space appears to remain, in which a nonzero ALP-photon coupling at low energies is compatible with asymptotic safety at high energies. However, this slice is thin compared to the systematic theoretical uncertainties that affect the location of the weak-gravity bound and the location of the region in which the ALP-photon coupling has an asymptotically free and a safe fixed point. Therefore, within the systematic uncertainties of our study, it is not possible to definitely say whether or not a nonzero ALP-photon coupling is compatible with asymptotic safety. Additional information to answer this question comes from previous studies, which determined the location of the gravitational fixed-point values. While affected by systematic uncertainties as well, they do not lie close to the region required for nonzero ALP-photon couplings. In summary, a fundamental pseudoscalar particle with an ALP-photon coupling thus appears to be disfavored by asymptotically safe quantum gravity. This also suggests that dark-matter candidates in asymptotic safety can probably not be fundamental ALPs, although the limitations of our study, in particular the simplicity of the matter sector, with just an ALP and an Abelian gauge field, should be kept in mind.

This result, while subject to systematic uncertainties, is another example of the predictive power of asymptotic safety. It is part of an accumulating body of evidence that the asymptotic safety paradigm is rather constraining, and the landscape of matter models at low energies which are compatible with asymptotic safety at high energies, could be rather restricted. In this way, the asymptotic-safety paradigm could become a valuable theoretical guide in the search for new physics beyond the Standard-Model and the understanding of fundamental questions such as the nature of the dark matter.\\

A main caveat of our study lie in our use of a truncation of the full dynamics. This results in the systematic uncertainties discussed above. Further, it implies that there might be further interacting fixed points that are generated by interactions beyond our truncation.

A second main caveat lies in our use of Euclidean spacetime signature, which is necessary to set up functional RG calculations with gravity. Thus, our conclusions might not extend to the ALP-gravity interaction in our universe, where spacetime has Lorentzian signature.\\

In the future, our study could be extended in several ways: First, it could be repeated in an extended truncation in order to reduce the systematic uncertainties. Second, it could be extended to settings where the ALP occurs as a Nambu-Goldstone boson of a spontaneously broken symmetry, such as for the QCD axion itself, as well as for dark-matter models with strong dynamics, see, e.g., \cite{Cai:2020bhd}.

\acknowledgments
This work is supported by a research grant (29405) from VILLUM FONDEN.  RRLdS thanks Nicolás Bernal for discussions on ULDM and axions.

\appendix 

\section{Beta functions and anomalous dimensions} \label{sec:betaApp}

\begin{figure}[!t]
	\hspace*{-3cm}\includegraphics[height=10cm]{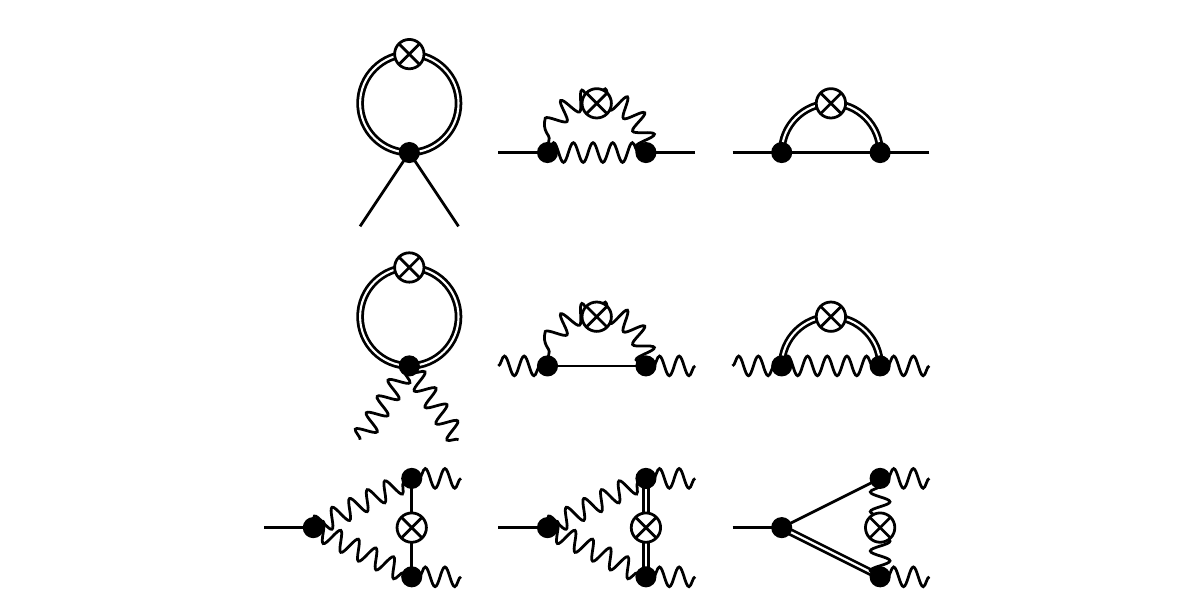}
	\caption{\label{fig:diagrams} 
	 From the diagrams in the first row, we can extract contributions to the anomalous dimension $\eta_\phi$ and to the beta function $\beta_{m^2}$. In the second row, we show the diagrams that contribute to the anomalous dimension $\eta_A$. In the third row, we show the diagrams that contribute to beta function $\beta_{g^2}$. All the contributions coming from the diagrams in the third row vanish for the gauge choice we are using throughout this work ($\alpha = 0 =\beta$). Straight lines represent the pseudo-scalar field. Wavy lines represent the photon field. Double straight lines represent the graviton field. These lines represent both external fields and propagators. Propagators connect the vertices and regulator insertions. The regulator insertion, denoted by a crossed circle, must appear on each of the internal lines in turn. For example, there are three versions of the last diagram, one where the regulator is inserted on the photon line, one where it is inserted on the scalar line, and one where it is inserted on the graviton line.}
\end{figure}

In this appendix, we provide additional details on the RG-equations that were used throughout this work. The beta functions obtained from the setup introduced in Sec.~\eqref{sec:settingALP+Grav} are
\begin{align}
	&\beta_{m^2} = (-2+\eta_\phi)\,m^2
	+ \frac{5\,(6-\eta_h)\,m^2\,G}{12\pi(1-2\Lambda)^2}
	+ \frac{(6-\eta_h)\,m^2\,G}{18\pi(1-4\Lambda/3)^2} \nonumber \label{eq:betam2Full}\\
	&\quad\quad -\frac{2\,(6-\eta_h)\, m^4\,G}{9\pi (1+m^2)(1-4\Lambda/3)^2}
	-\frac{2\,(6-\eta_\phi)\, m^4\,G}{9\pi (1+m^2)^2(1-4\Lambda/3)} \,, \\
	&\beta_{g^2}= \left(2 + 2 \eta _A+\eta _{\phi }\right) g^2 \,, \label{eq:betag2Full}
\end{align}
with anomalous dimensions $\eta_\phi$ and $\eta_A$ given by
\begin{align}
	&\eta_\phi = \frac{(8-\eta_A)\,g^2}{128\pi^2} 
	+ \frac{(8-\eta_h)\,G}{144\pi (1+m^2)(1-4\Lambda/3)^2}
	+ \frac{2\,(6-\eta_h) \,m^2 \,G}{9\pi (1+m^2)(1-4\Lambda/3)^2} \nonumber \\
	&\qquad+\frac{(8-\eta_\phi)\,G}{144\pi (1+m^2)^2(1-4\Lambda/3)}
	- \frac{4 \,m^4\,G}{3\pi (1+m^2)^2(1-4\Lambda/3)^2} \,, \label{eq:etaphiFull} \\
	&\eta_A = \frac{(8-\eta_A)\,g^2}{384\pi^2(1+m^2)}
	+ \frac{(8-\eta_\phi)\,g^2}{384\pi^2(1+m^2)^2}  \nonumber \\
	&\qquad + \frac{5\,(6-\eta_h)\,G}{18\pi (1-2\Lambda)^2} 
	- \frac{5\,(8-\eta_h)\,G}{36\pi(1-2\Lambda)^2} 
	-\frac{5\,(8-\eta_A)\,G}{36\pi(1-2\Lambda)} \, \label{eq:etaAFull} .
\end{align}
The beta functions $\beta_{m^2}$ and $\beta_{g^2}$ and the anomalous dimensions $\eta_\phi$ and $\eta_A$ receive contributions from the diagrams in Fig.~\eqref{fig:diagrams}.
We can also compute the graviton anomalous dimension $\eta_h$, but its explicit form is not relevant for the discussion presented here. 

In this work, we adopted an approximation that is defined by neglecting the anomalous dimensions arising from regulator insertions $\partial_t \textbf{R}_k$, but keeping the anomalous dimensions arsing from the canonical part of the beta functions $\beta_{m^2}$ and $\beta_{g^2}$. This is equivalent to setting $\eta_h = 0$ everywhere and $\eta_\phi = 0 = \eta_A$ on the right hand side of \eqref{eq:etaphiFull} and \eqref{eq:etaAFull}. In this approximation, we also set $\eta_\phi = 0$ in the last term of \eqref{eq:betam2Full}. This approximation is justified by the fact that the anomalous dimensions coming from $\partial_t \textbf{R}_k$ appear with the particular combinations $(6 - \eta_i)$ and $(8 - \eta_i)$. In the (near-)perturbative regime we expect that the anomalous dimensions should not be large and, therefore, they should be suppressed when appearing in the form $(6 - \eta_i)$ or $(8 - \eta_i)$.

\section{Pure-matter case}\label{sec:purematter}

In the absence of gravity, i.e., for $G=0$, the only viable fixed point for $m^2$ and $g^2$ is the free fixed point.
Therefore, the ALP-model is not asymptotically safe on its own. This result is a consequence of the beta functions
\begin{align}
	&\beta_{m^2}= (-2 + \eta_\phi)m^2 \,, \label{eq:betam2_pureMatter} \\
	&\beta_{g^2}= \left(2 + 2 \eta _A+\eta _{\phi }\right) g^2 \,, \label{eq:betag2_pureMatter} 
\end{align}
with anomalous dimensions
\begin{align}
	&\eta_\phi = \frac{(8-\eta_A)\,g^2}{128\pi^2} \,, \label{eq:etaphi_pureMatter} \\
	&\eta_A = \frac{(8-\eta_A)\,g^2}{384\pi^2(1+m^2)}
	+ \frac{(8-\eta_\phi)\,g^2}{384\pi^2(1+m^2)^2}   \,. \label{eq:etaA_pureMatter} 
\end{align}
This set of RG-equations can be obtained by setting $G=0$ in \eqref{eq:betam2Full}, \eqref{eq:betag2Full}, \eqref{eq:etaphiFull} and \eqref{eq:etaAFull}. The RG-flow of the ALP-photon system (in the absence of gravity) was first computed in \cite{Eichhorn:2012uv}. Our result for the scalar anomalous dimension $\eta_\phi$ is slightly different from the one reported in \cite{Eichhorn:2012uv} due to a different gauge choice.

Using the approximation adopted throughout this work, i.e., neglecting the anomalous dimension arising from the regulator insertion $\partial_t \textbf{R}_k$ (see App. \ref{sec:betaApp}), the beta functions for $m^2$ and $g^2$ can be written as
\begin{align}
	&\beta_{m^2} = -2 m^2 + \frac{m^2\,g^2}{16\pi^2} \,, \label{eq:betam2LO}\\
	&\beta_{g^2} = 2 g^2 + \frac{(7+8\,m^2+3\,m^4) \,g^4}{48\pi^2 (1+m^2)^2} \,. \label{eq:betag2LO}
\end{align}
From Eq. \eqref{eq:betag2LO}, we can see that the coefficient of $g^4$ is strictly positive. Thus, we cannot obtain an interacting fixed point with real values for $g$. Therefore, within the current approximation, the only possibility of fixed points with non-negative values for $m^2$ and $g^2$ is the free fixed point $(m_*^2,g_*^2) = (0,0)$.
	
This conclusion can be extended beyond the approximation where we neglect the anomalous dimensions coming from regulator insertion. Solving \eqref{eq:etaphi_pureMatter} and \eqref{eq:etaA_pureMatter} for $\eta_\phi$ and $\eta_A$ and plugging the result into \eqref{eq:betam2_pureMatter} and \eqref{eq:betag2_pureMatter}, we find
\begin{align}
	&\beta_{m^2} = -2m^2 + 
	\frac{8\,(384\pi^2 (1+m^2)^2 - g^2 )\,m^2g^2}{49152 \pi^4 (1+m^2)^2 + 128 \pi^2 (1+m^2)\,g^2 - g^4} \, , \label{eq:betam2_pureMatterFull}\\
	&\beta_{g^2} = 2g^2 + 
	\frac{8\,\big(128\pi^2 (7+8\,m^2+3\,m^4) - 3g^2 \big)\,g^4}{49152 \pi^4 (1+m^2)^2 + 128 \pi^2 (1+m^2)\,g^2 - g^4}	 \,. \label{eq:betag2_pureMatterFull}
\end{align}
Solving the fixed-point equations $\beta_{m^2} = 0$ and $\beta_{g^2}=0$, we find a fixed-point candidate with non-negative values for $m^2$ and $g^2$: $(m_*^2,g_*^2) = (0,2943.28)$. However, we can argue that this fixed-point candidate is not reliable due to its critical exponents, $(\theta_1,\theta_2)_\textmd{int.} \approx (-1498,129)$, that deviates considerably from the canonical values $(\theta_1,\theta_2)_\textmd{free} = (-2,2)$ associated with the free fixed point. 

We briefly point out a difference between our result in comparison with \cite{Eichhorn:2012uv}, which uses a different gauge-fixing in the abelian gauge sector. Ref.~\cite{Eichhorn:2012uv} points out the existence of an exceptional RG-trajectory that avoids Landau poles, despite not being asymptotically safe. This exceptional trajectory is defined by an exceptional point $(m^2_\textmd{exc},g^2_\textmd{exc})$ where the divergence coming from a zero on the numerator of the beta functions $\beta_{m^2}$ and $\beta_{g^2}$ is canceled by a zero on the numerators. We cannot find such exceptional point from the beta functions \eqref{eq:betam2_pureMatterFull} and \eqref{eq:betag2_pureMatterFull} and, therefore, we conclude that this trajectory is not present in our setup. If such a trajectory was indeed present as a physical scenario, we would expect it to be present independently of the choice of gauge.

\bibliography{references}
\end{document}